# Electrical switching in a magnetically intercalated transition metal dichalcogenide

Nityan L. Nair[1,2,†], Eran Maniv[1,2,†], Caolan John[1,2], Spencer Doyle[1,2], J. Orenstein[1,2], James G. Analytis[1,2,*]

[1]*Department of Physics, University of California, Berkeley, California 94720, USA*
[2]*Materials Science Division, Lawrence Berkeley National Laboratory, Berkeley, California 94720, USA*
*†These authors contributed equally to this work*
*\*Corresponding author*

**Recent advances in tuning the correlated behavior of graphene and transition-metal dichalcogenides (TMDs) have opened a new frontier in the study of many-body physics in two dimensions and promise exciting possibilities for new quantum technologies. An emerging field where these materials have yet to make a deep impact is the study of antiferromagnetic (AFM) spintronics – a relatively new research direction that promises technologies that are insensitive to external magnetic fields, fast switching times, and reduced crosstalk[1–3]. In this study we present measurements on the intercalated TMD $Fe_{1/3}NbS_2$ which exhibits antiferromagnetic ordering below 42K[4,5]. We find that current densities on the order of $10^4$ A/cm$^2$ can reorient the magnetic order, the response of which can be detected in the sample's resistance. This demonstrates that $Fe_{1/3}NbS_2$ can be used as an antiferromagnetic switch with electronic "write-in" and "read-out". This switching is found to be stable over time and remarkably robust to external magnetic fields. $Fe_{1/3}NbS_2$ is a rare example of an AFM system that exhibits fully electronic switching behavior in single crystal form, making it appealing for low-power, low-temperature memory storage applications. Moreover, $Fe_{1/3}NbS_2$ is part of a much larger family of magnetically intercalated TMDs, some of which may exhibit the switching behavior at higher temperatures and form a platform from which to build tunable AFM spintronic devices[6,7].**

AFM memory storage devices have been long sought-after in the field of spintronics. Compared to the their widely used ferromagnetic (FM) counterparts, AFM memory promises several key improvements. AFMs do not produce external stray fields, making memory stored in these devices invisible to external magnetic probes and allowing individual devices to be more tightly packed on-chip[1,2]. They possess ultrafast spin dynamics; AFM devices have been recently demonstrated to switch at THz speeds, significantly faster than their GHz-limited FM counterparts[3,8]. Finally, they couple weakly to external magnetic fields, making AFM devices robust to magnetic perturbations. Combined, these properties make AFMs appealing for high-density, ultrafast, extremely stable memory storage applications. Their insensitivity to field, however, makes manipulating and detecting AFMs difficult, limiting their widespread adoption primarily to passive layers in FM heterostructure devices[9,10]. Only two examples have emerged in which the AFM order can be demonstrably manipulated and detected by applied currents: $CuMnAs$ and $Mn_2Au$[11,12].

In this work, we show that certain magnetically intercalated TMDs, known for their diverse and highly tunable electronic and magnetic properties, can possess switchable AFM orders. We find that electrical current densities on the order of $10^4$ A/cm$^2$, more than two orders of magnitude lower than CuMnAs and Mn$_2$Au, can be used to switch the AFM order in Fe$_{1/3}$NbS$_2$ between two stable configurations through the intrinsic magneto-electric coupling of the crystal, making it appealing for low-power applications. The anisotropic magnetoresistance (AMR) allows the orientation of the AFM to be determined from simple resistivity measurements. In this manner, we demonstrate the use of Fe$_{1/3}$NbS$_2$ as a low-power magnetic memory bit with electronic write-in and read-out. Although Fe$_{1/3}$NbS$_2$ only exhibits switching behavior below its AFM ordering temperature of 42K, it is part of a larger class of magnetically intercalated TMDs that is just beginning to be explored. Within this class there exist compounds with higher ordering temperatures that may extend the switching behavior of Fe$_{1/3}$NbS$_2$ to room temperature and allow for application-specific tuning of device properties[13–15]. This paves the way for using Fe$_{1/3}$NbS$_2$, and perhaps magnetically intercalated TMDs more generally, in a new generation of electrically switchable AFM spintronic devices.

Fe$_{1/3}$NbS$_2$ consists of iron atoms intercalated in a periodic lattice between planes of the 2H phase of NbS$_2$. Single crystals of Fe$_{1/3}$NbS$_2$ were grown via vapor transport and fabricated into transport devices using focused ion beam (FIB) techniques (see supplement)[16,17]. A typical eight-contact device is shown in Figure 1a. The transverse resistance of the sample can be measured by driving a 100μA AC current across one leg of the device and measuring the resulting voltage produced across the orthogonal leg (yellow contacts) using standard lock-in measurement techniques. In tandem, DC current pulses can be applied along two orthogonal directions (red and blue contacts). As shown in Figure 1b, the application of DC current pulses at low temperature serves to switch the sample between two magnetic states, which can then be read out via the transverse resistivity. A $5.4 \times 10^4$ A/cm$^2$ current applied for 10ms along the blue bar switches the device into a low transverse resistance state. Applying the sample pulse along the red bar switches the device into a high transverse resistance state. This switching behavior is both repeatable and stable. The device can be repeatably switched from one resistance state into the other with subsequent application of a current pulse along the two orthogonal directions. Applying additional current pulses along the same axis or changing the sign of the current does not change the resistance state of the sample, indicating that a *single* pulse saturates the response (see supplement). Moreover, once in the high or low resistance state the device is stable and the measured transverse resistance does not decay over the 30-second intervals shown in Figure 1b.

The switching behavior of Fe$_{1/3}$NbS$_2$ originates from the coupling of the charge current to the magnetic order. Figures 1c-d show out-of-plane and in-plane magnetization measurements performed on a bulk single crystal, which exhibit an anomaly at the AFM ordering temperature of 42K. Figure 2a shows the temperature dependence of the switching, which perfectly correlates with this transition. The switching amplitude is extracted by fitting a square wave to R$_\perp$ and decreases as the temperature increases, up to approximately 40K, above which switching is no longer observable (see inset). Application of an out-of-plane magnetic field (Figure 2b) also suppresses the switching behavior, although it can still be observed at fields as high as 12T. This insensitivity to magnetic field is a strong indication that the switching stems from an AFM order and makes $Fe_{1/3}NbS_2$ and related compounds suitable candidates for robust, low-temperature memory storage applications.

Rotating the device geometry, as shown in Figure 3a, can help elucidate the mechanism of the switching behavior. The AC excitation current and the associated four-point resistance measurements ($R_∥$ and $R_⊥$) were rotated between contacts while keeping the direction of the DC current pulses fixed. With the AC current 45° from the DC pulses (column 1), the same configuration measured in Figures 1 and 2, all the switching is observed in the transverse channel ($R_⊥$) while almost no signal is observed in the parallel resistance ($R_∥$). Rotating the contacts to 90° (column 2) reverses this. The switching appears parallel to the current while the signal in the transverse channel is suppressed. Rotating the contacts further, to 135° and 180°, moves the switching signal back into the perpendicular and parallel channels, respectively, however the sign is now reversed. Namely, application of the same pulse sequence results in a reversed ordering of high and low resistance states.

The angle dependence of the switching is intimately connected to the AMR. The AMR was measured by field cooling a bulk crystal in a 9 Tesla in-plane field from above the AFM transition to 2K. The magnetic field was then turned off, and $R_∥$ and $R_⊥$ were measured. This process was repeated at several angles, as shown in Figure 3b. After background subtraction, $R_∥$ and $R_⊥$ show the characteristic sinusoidal AMR behavior offset by 45° [18]. The amplitude of the AMR signal is completely suppressed by 40K (Figure 3b, inset), the same temperature at which the switching behavior disappears and close to the AFM transition.

There are two important things to note in the AMR data. First, the AMR presented in Figure 3b is the *zero-field* AMR. This differs from a conventional AMR measurement in which $R_∥$ and $R_⊥$ are measured in the presence of a finite magnetic field. (Similar field-induced memory effects stemming from a zero-field AMR have been observed in the AFM systems FeRh and MnTe[19,20]). The fact that this quantity is non-zero means that the in-plane field influences the direction of the AFM order, which remains frozen-in even after the field has been removed. Like the current pulses $\vec{J}$, the magnetic field $\vec{B}$ can be used to write information into $Fe_{1/3}NbS_2$, which can then be read out via the AMR after cooling.

Second, the zero-field AMR also has the same angle dependence as the switching amplitude. The AMR is independent of the sign of $\vec{B}$, as is the switching amplitude of the sign of $\vec{J}$. Where $R_∥$ or $R_⊥$ in the AMR is maximized, the switching amplitude is maximized in the same channel. Where the $R_∥$ or $R_⊥$ go through zero, the switching behavior is suppressed. Where $R_∥$ or $R_⊥$ become negative, the switching order inverts. This implies that the application of a current pulse to the device is similar to an in-plane magnetic field: they both "write" a preferred orientation into the AFM state, which can then be electronically read out via the AMR. In other words, both a current pulse and an applied magnetic field in the same direction act to rotate the principal axes of the resistivity tensor in the same way. In this manner, the device can be used to store information with electrical write-in and read-out.

Figure 4 shows the dependence of the switching behavior on the current density (a) and duration (b) of the DC pulses. At large current densities and durations, the switching amplitude saturates, indicating that the in-plane magnetic order has been fully polarized by the DC pulse. At low densities/durations, however, the switching amplitude is non-monotonic, exhibiting a local maximum. Moreover, although the device appears to exhibit switching in this region, the amplitude is smaller than in the saturated region and the high and low resistance states are not perfectly repeatable; the resistance values appear to change between pulse sets. This may be due to domains of the in-plane AFM order that cannot be fully polarized below a critical threshold. Although not fully saturated, switching can be observed at current densities as low as $2.7 \times 10^4$ A/cm$^2$ and pulse durations as short as 10μs (the limit of our experimental apparatus), both orders of magnitude lower that what has been previously reported for DC pulses in CuMnAs and Mn$_2$Au[11,12].

The magnetic order of Fe$_{1/3}$NbS$_2$ appears to be more complicated than previous measurements have suggested, which determined the low-temperature AFM order to be aligned parallel to the c-axis in a wurtzite-type geometry[4]. An applied in-plane magnetic field would be expected to couple isotropically to such an order. In contrast, our AMR and magnetization measurements indicate that some component of the AFM order lies in the a-b plane. This could be due to canting of the spins or the coexistence of a more complicated helical magnetic order, as has been observed in other intercalated TMDs such as Cr$_{1/3}$NbS$_2$[21]. Regardless, this in-plane component must be the origin of both the zero-field AMR and as discussed below, the switching behavior.

In order to understand the switching behavior, it is useful to treat it as two separate processes: "write-in" in which information is encoded into the AFM state via the current pulses, and "read-out" in which the AFM state is probed by the resistivity measurement. The read-out process is equivalent to the aforementioned AMR measurements. The anisotropic scattering of conduction electrons from the localized iron moments changes the measured resistivities depending on the relative orientation of the AFM order and the AC probe current[18,22]. In this manner, the resistivity tensor is sensitive to the direction of the in-plane AFM order. When the in-plane component of the spins is rotated 90° by a current pulse, the measured resistivity switches from a maximum to a minimum (or vice versa), reading-out the information encoded into the magnetic state.

In the write-in process, the current pulses reorient the Néel vector and hence rotate the principal axes of the AMR-derived resistivity tensor. Importantly, the zero-field AMR measurements (Figure 3), show that the underlying mechanism must couple the applied magnetic fields and electric currents to the AFM order in the same way. The lack of inversion symmetry in Fe$_{1/3}$NbS$_2$ gives an important clue as to how this can arise. A current pulse in $Fe_{1/3}NbS_2$, will in general be spin polarized due to the inverse spin galvanic effect (ISGE). The ISGE takes advantage of the Rashba spin-orbit interaction, which is allowed in non-centrosymmetric systems, producing asymmetric scattering between non-degenerate spin subbands in the presence of an electric field[23–25]. This spin polarized current will then exert a spin transfer torque (STT) on the Néel order. In the most general case, the STT is given by

$$\tau \sim \vec{M}^{A/B} \times \vec{p} + \vec{M}^{A/B} \times (\vec{M}^{A/B} \times \vec{p})$$

where $\vec{M}^{A/B}$ is the local magnetic moment on the $A/B^{th}$ sublattice and $\vec{p}$ is the net spin polarization of the injected charge current[26–28]. The first term in the STT is referred to as field-like and the second as antidamping-like. Because the STT is of the form $\tau = M \times B_{eff}$, an effective magnetic field can be extracted. $B_{eff}^F = \vec{p}$ for the field-like term and $B_{eff}^{AD} = \vec{M}^{A/B} \times \vec{p}$ for the antidamping-like term. In a real system in which the moments may relax angular momentum to the lattice, one expects that spins will tend to align along the local effective field. Which of these effective field acts parallel to the write current $\vec{J}$ depends on the direction of $\vec{p}$.

In a conventional heterostructure device, a nonzero spin polarization, $\vec{p}$, is achieved by injecting charge carriers through a fixed FM layer and is parallel to the FM moment[29]. In the present case, the ISGE and the broken inversion symmetry of the lattice lead to a net spin polarization given by $\vec{p} = \hat{z} \times \vec{J}$, where $\hat{z}$ is the unit vector along the direction of the relevant broken mirror symmetry, in this case the c-axis[30]. Given that the spin polarization is primarily oriented along the c-axis, this leads both field-like and antidamping-like effective fields to lie in-plane. However, only the anti-damping term has $B_{eff}^{AD}$ parallel to the applied current pulse $\vec{J}$, so that it will locally transform in the same way as the applied magnetic field, and therefore lead to an angle dependence in the switching that mirrors that of the AMR, as is observed. This mechanism is discussed in detail in references [29,30].

The determination that the antidamping-like term of the STT is the primary driver of the switching is supported by symmetry considerations as well. For a constant spin polarization $\vec{p}$ the antidamping effective field will be staggered with the sublattice periodicity. Therefore, we expect it to couple strongly to the AFM order since it inherits the same translational symmetry. The field-like effective field, on the other hand, will not be staggered and therefore would not be expected to couple as efficiently. In fact, these symmetry considerations lead the field-like term to couple more strongly to FM systems, and STT-driven switching as a result of this coupling has been observed in FM heterostructures[31,32].

This mechanism is in contrast to CuMnAs and $Mn_2Au$, whose inversion symmetry implies the IGSE must produce a staggered spin polarization and therefore the field-like component of the STT is thought to couple more strongly to the AFM order[11,12]. To our knowledge, $Fe_{1/3}NbS_2$ is the only AFM system in which the antidamping-like term drives the switching and this difference may be responsible for the enhanced response.

We have demonstrated that current pulses can be used to rotate the in-plane component of the Néel vector in $Fe_{1/3}NbS_2$ via the antidamping-like STT, thereby electronically writing information into the AFM state. This information can subsequently be read-out with electronic resistivity measurements via the zero-field AMR, forming an electronically-accessible AFM memory device. The mechanism behind this switching appears to be unique to $Fe_{1/3}NbS_2$ and may be responsible for the low current densities required and rapid saturation of the response, making this material an excellent candidate for low-temperature spintronics applications. More exciting, however, is the fact that $Fe_{1/3}NbS_2$ is just one of numerous compounds in the family of intercalated TMDs[6,7,13–15]. The demonstration of AFM switching in $Fe_{1/3}NbS_2$, opens the door to search for its presence in the hundreds of other related van der Waals materials, potentially realizing the phenomenon at room temperature and allowing for the application-specific tuning of device properties.

**Acknowledgements:**

This work was supported as part of the Center for Novel Pathways to Quantum Coherence in Materials, an Energy Frontier Research Center funded by the U.S. Department of Energy, Office of Science, Basic Energy Sciences. J.G.A. and N.L.N. received support from the Gordon and Betty Moore Foundation's EPiQS Initiative through Grant No. GBMF4374. J.O. received support from the Gordon and Betty Moore Foundation's EPiQS Initiative through Grant GBMF4537. FIB device fabrication was performed at the National Center for Electron Microscopy (NCEM) at the Molecular Foundry. Work at the Molecular Foundry was supported by the Office of Science, Office of Basic Energy Sciences, of the U.S. Department of Energy under Contract No. DE-AC02-05CH11231.

**Author Contributions:**

J.G.A. and E.M. conceptualized the experiment. S.D. and C.J. performed crystal synthesis and magnetization measurements. N.L.N. fabricated FIB microstructure devices. N.L.N. and E.M. conducted transport measurements and data analysis. N.L.N. wrote the manuscript with input from all coauthors.

**Competing Interests:**

The authors declare no competing financial interests.

**References:**

1. Gomonay, O., Jungwirth, T. & Sinova, J. Concepts of antiferromagnetic spintronics. *Phys. status solidi - Rapid Res. Lett.* **11,** 1700022 (2017).

2. Baltz, V. *et al.* Antiferromagnetic spintronics. *Rev. Mod. Phys.* **90,** 015005 (2018).

3. Olejník, K. *et al.* Terahertz electrical writing speed in an antiferromagnetic memory. *Sci. Adv.* **4,** eaar3566 (2018).

4. Van Laar, B., Rietveld, H. M. & Ijdo, D. J. W. Magnetic and crystallographic structures of MexNbS2 and MexTaS2. *J. Solid State Chem.* **3,** 154–160 (1971).

5. Gorochov, O., Blanc-soreau, A. Le, Rouxel, J., Imbert, P. & Jehanno, G. Transport properties, magnetic susceptibility and Mössbauer spectroscopy of Fe 0.25 NbS 2 and Fe 0.33 NbS 2. *Philos. Mag. B* **43,** 621–634 (1981).

6. Manzeli, S., Ovchinnikov, D., Pasquier, D., Yazyev, O. V & Kis, A. 2D transition metal dichalcogenides. *Nat. Rev. Mater.* **2,** 17033 (2017).

7. Chhowalla, M. *et al.* The chemistry of two-dimensional layered transition metal dichalcogenide nanosheets. *Nat. Chem.* **5,** 263–275 (2013).

8. Garello, K. *et al.* Ultrafast magnetization switching by spin-orbit torques. *Appl. Phys. Lett.* **105,** 212402 (2014).

9. Jungwirth, T., Marti, X., Wadley, P. & Wunderlich, J. Antiferromagnetic spintronics. *Nat. Nanotechnol.* **11,** 231–241 (2016).

10. Železný, J., Wadley, P., Olejník, K., Hoffmann, A. & Ohno, H. Spin transport and spin torque in antiferromagnetic devices. *Nat. Phys.* **14,** 220–228 (2018).

11. Wadley, P. *et al.* Electrical switching of an antiferromagnet. *Science (80-. ).* **351,** 587–590 (2016).

12. Bodnar, S. Y. *et al.* Writing and reading antiferromagnetic Mn2Au by Néel spin-orbit torques and large anisotropic magnetoresistance. *Nat. Commun.* **9,** 348 (2018).

13. Parkin, S. S. P. & Friend, R. H. 3 d transition-metal intercalates of the niobium and tantalum dichalcogenides. I. Magnetic properties. *Philos. Mag. B* **41,** 65–93 (1980).

14. Parkin, S. S. P. & Friend, R. H. 3 d transition-metal intercalates of the niobium and tantalum dichalcogenides. II. Transport properties. *Philos. Mag. B* **41,** 95–112 (1980).

15. Friend, R. H., Beal, A. R. & Yoffe, A. D. Electrical and magnetic properties of some first row transition metal intercalates of niobium disulphide. *Philos. Mag. A J. Theor. Exp. Appl. Phys.* **35,** 1269–1287 (1977).

16. Moll, P. J. W. *et al.* High magnetic-field scales and critical currents in SmFeAs(O, F) crystals. *Nat. Mater.* **9,** 628–633 (2010).

17. Moll, P. J. W. Focused Ion Beam Microstructuring of Quantum Matter. *Annu. Rev. Condens. Matter Phys* **9,** 147–162 (2018).

18. McGuire, T. & Potter, R. Anisotropic magnetoresistance in ferromagnetic 3d alloys. *IEEE Trans. Magn.* **11,** 1018–1038 (1975).

19. Marti, X. *et al.* Room-temperature antiferromagnetic memory resistor. *Nat. Mater.* **13,** 367–374 (2014).

20. Kriegner, D. *et al.* Multiple-stable anisotropic magnetoresistance memory in antiferromagnetic MnTe. *Nat. Commun.* **7,** 11623 (2016).

21. Togawa, Y. *et al.* Chiral Magnetic Soliton Lattice on a Chiral Helimagnet. *Phys. Rev. Lett.* **108,** 107202 (2012).

22. Park, B. G. *et al.* A spin-valve-like magnetoresistance of an antiferromagnet-based tunnel junction. *Nat. Mater.* **10,** 347–351 (2011).

23. Silov, A. Y. *et al.* Current-induced spin polarization at a single heterojunction. *Appl. Phys. Lett.* **85,** 5929–5931 (2004).

24. Kato, Y. K., Myers, R. C., Gossard, A. C. & Awschalom, D. D. Current-Induced Spin Polarization in Strained Semiconductors. *Phys. Rev. Lett.* **93,** 176601 (2004).

25. Ganichev, S. D. *et al.* Spin-galvanic effect. *Nature* **417,** 153–156 (2002).

26. Ralph, D. C. & Stiles, M. D. Spin transfer torques. *J. Magn. Magn. Mater.* **320,** 1190–1216 (2008).

27. Hellman, F. *et al.* Interface-induced phenomena in magnetism. *Rev. Mod. Phys.* **89,** 025006 (2017).

28. Gomonay, O., Baltz, V., Brataas, A. & Tserkovnyak, Y. Antiferromagnetic spin textures and dynamics. *Nat. Phys.* (2018). doi:10.1038/s41567-018-0049-4

29. Gomonay, H. V & Loktev, V. M. Spin transfer and current-induced switching in antiferromagnets. *Phys. Rev. B* **81,** 144427 (2010).

30. Železný, J. *et al.* Relativistic Néel-Order Fields Induced by Electrical Current in Antiferromagnets. *Phys. Rev. Lett.* **113,** 157201 (2014).

31. Maekawa, S. (Sadamichi). *Concepts in Spin Electronics*. (Oxford University Press, 2006). doi:10.1093/acprof:oso/9780198568216.001.0001

32. Chappert, C., Fert, A. & Van Dau, F. N. The emergence of spin electronics in data storage. *Nat. Mater.* **6,** 813–823 (2007).

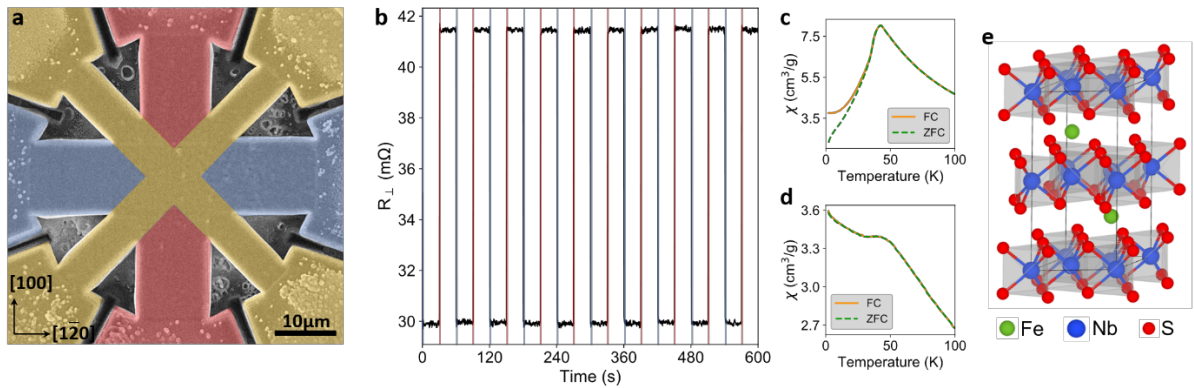

**Figure 1: Electrical switching of Fe$_{1/3}$NbS$_2$**

a) A false-color SEM image of a Fe$_{1/3}$NbS$_2$ switching device. The transverse resistivity ($R_\perp$) is measured using the yellow contacts. A 100μA AC probe current is applied along one yellow bar, while the voltage drop is measured along the orthogonal bar using standard lock-in techniques. Simultaneously, DC current pulses can be applied along the red and blue contacts in the [100] and [1$\bar{2}$0] directions.
b) When orthogonal current pulses are applied, the transverse resistivity switches between two states. Applying $5.4 \times 10^4$ A/cm$^2$ for 10ms along the blue contacts switches the device into a low transverse resistivity state. Applying the same pulse along the red contacts switches the device into a high state. The time between pulses is 30 seconds and the switching has been repeated 10 times to show the robustness of this behavior.
c) C-axis magnetization measurements show a peak at 42K in both the field cooled and zero field cooled traces indicating the presence of an AFM transition.
d) In-plane magnetization measurements also show a weak peak at 42K.
e) The crystal structure of Fe$_{1/3}$NbS$_2$ is that of 2H-NbS$_2$ with iron atoms intercalated between layers. At this stoichiometry, the iron intercalants form an ordered lattice with space group P6$_3$22 (no. 182).

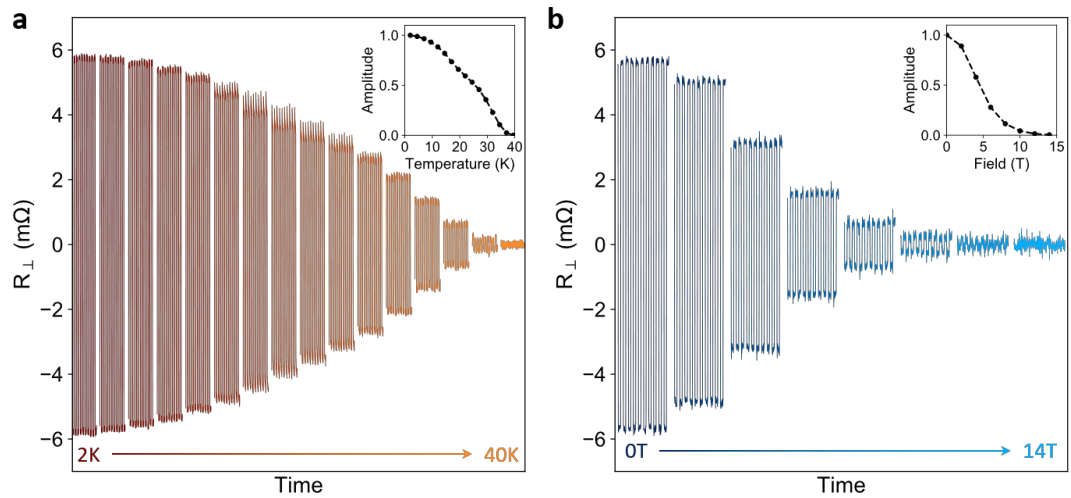

**Figure 2: Temperature and field dependence**

a) The switching behavior is suppressed by temperature. By 40K, the switching amplitude is completely suppressed, as show in the inset. The temperature-dependent background of $R_\perp$ has been subtracted from all curves to highlight only the switching component.

b) The switching behavior at 2K is suppressed by magnetic field, although it shows surprising robustness and can be observed at fields as high at 12T. The magnetic field is applied along the c-axis. The field-dependent background of $R_\perp$ has been subtracted to highlight only the switching component. The extracted switching amplitude is shown in the inset.

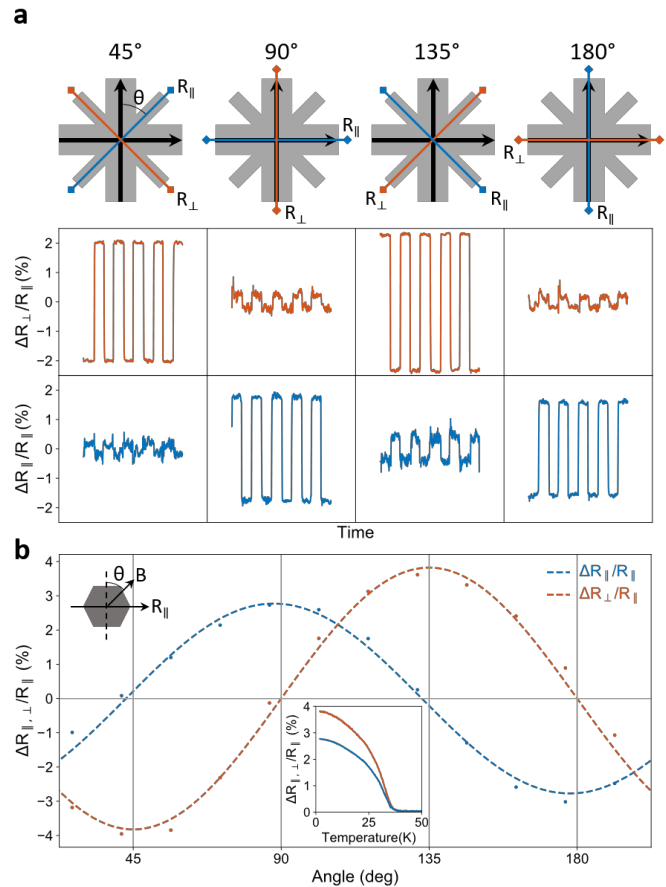

**Figure 3: Geometry dependence of the switching and correlation to AMR**

a) The switching behavior is dependent on device geometry. As the angle between the AC probe current and the DC write pulses is rotated (top row), the switching signal moves between the transverse (red, middle row) and longitudinal (blue, bottom row) resistivity channels, picking up a sign change between 90° and 135°. Black arrows denote the fixed directions of the DC pulses, with the horizontal bar pulsed first followed by the vertical bar, repeated five times. Red denotes the transverse resistivity. Blue denotes the longitudinal resistivity and the direction of the AC probe current. The measurement configuration at 45° is equivalent to that in Figure 1.
b) The zero-field AMR shows a very similar angle dependence. Every 45° rotation shifts the signal from one resistivity channel into the other. Moreover, the sign of the AMR switches in the same angular range as the sign change in the switching. As shown in the inset, the AMR vanishes above approximately 38K, the same temperature at which the switching behavior disappears. A constant background has been subtracted from both curves.

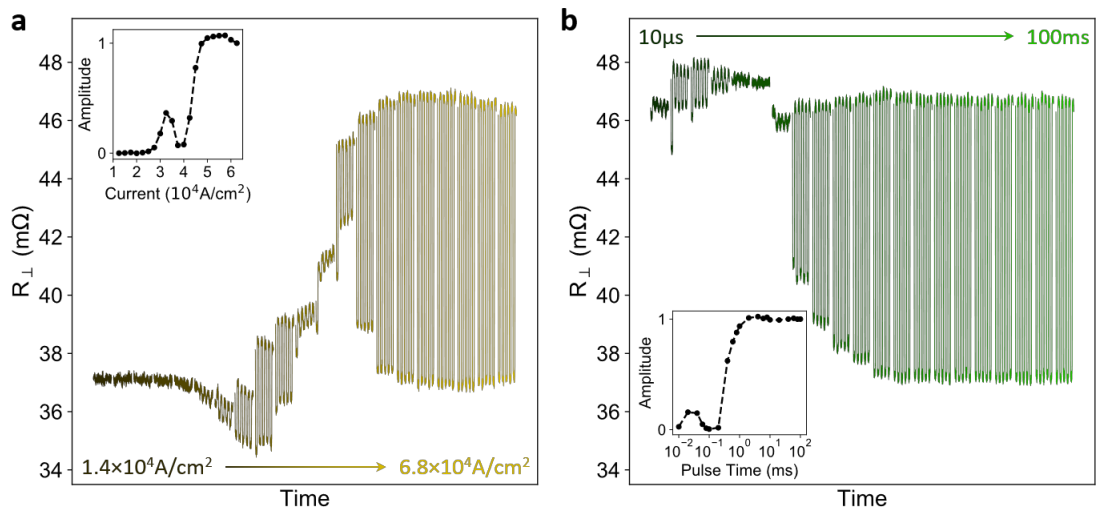

**Figure 4: Dependence on current density and duration**

a) The switching amplitude saturates at large current densities but shows non-monotonicity and a local maximum at small currents. Switching can be observed at current densities as low as $2.7 \times 10^4 A/cm^2$. The extracted switching amplitude is plotted in the inset. Measurements were performed at 2K in the absence of an external field with a 20ms pulse duration.

b) The pulse duration shows a very similar behavior to the current dependence, with a local maximum followed by saturation of the switching amplitude. Switching is observed as low as 10μs, the limit of our experimental apparatus. Measurements were performed at 2K in the absence of an external field with a $5.4 \times 10^4$ A/cm² current density.